\documentclass[oneside,12pt]{article}

\title{New Physics and Neutrino Oscillation}
\author{M.~Ochman, R.~Szafron and  M.~Zralek\\
 {\small Institute of Physics, University of Silesia, Poland}
}
\date{}

\begin{document}

\maketitle

\begin{abstract}  Description of neutrino oscillation in the case of Non-Standard neutrino Interaction (NSI) is briefly presented. The NSI causes the entanglement between internal degrees of freedom of neutrinos (mass, spin, flavour) and other accompanying particles in the production and detection processes. In such case neutrinos are mostly in the mixed states. Role of the density matrix in description of neutrino oscillation process is shortly explained.
\end{abstract}

\vspace{1cm}

Theory of neutrino oscillation found in late sixties \cite{Pontecorvo} works well and is almost commonly accepted.  Now days two methods describing particles oscillation are considered, one based on the quantum mechanics (QM) and the second which use the quantum field theory (QFT) (for a review see \cite{Theory of neutrino oscillation}). For the standard relativistic neutrino interaction both methods are equivalent, and give the same result for final neutrino oscillation rate \cite{Akhmedov and Kopp}.  In the case of a NSI there is no, up to now, uniform way of describing the neutrino oscillation process. In the standard approach the state of produced neutrinos does not depend on the  state of accompanying particles, states are disentangled. In such a case neutrinos oscillate  in the  way  independent of the spins arrangement  of the other particles present in the production and detection process. In the QFT approach, and in most studies of the neutrino oscillation  beyond the Standard Model (SM) \cite{Grossman}, oscillation process is assumed to be  universal, where probability of  neutrino transition is possible to be defined and depends only on the initial and final states.

However,  beyond the SM the production and detection neutrino states may differ from states which are used to construct the Lagrangian of the model before the spontaneous symmetry breaking, the so called flavour states \cite{Papers on oscillation in NSI}.  It can happen also that the right-handed neutrino fields appear, although there are  a fairly strong limits on couplings involving these fields \cite{bounds on RHN}. This causes that neutrino states in the production, detection and propagation processes need not to be given by the simple combination of neutrino mass states with the same helicity,
\begin{equation} 
\vert\nu_{\alpha},\downarrow\rangle =\sum_{i}U_{\alpha,i}^{*}\vert\nu_{i},\downarrow\rangle,\ \ \   \label{neutrin SM states} \vert\overline{\nu}_{\alpha},\uparrow\rangle =\sum_{i}U_{\alpha,i}\vert\overline{\nu}_{i},\uparrow\rangle,
\end{equation}
which is the basic assumption of QM description of the standard neutrino oscillation. 
In most approaches beyond the SM,  instead of  the states (\ref{neutrin SM states}) the new one, separate for production (p) and detection (d) are defined
\begin{equation}
\label{new production and detection states}
\vert\nu_{\alpha}^{p}\rangle =\sum_{i}U_{\alpha,i}^{p}\vert\nu_{i}\rangle,  \ \ \   \vert\nu_{\beta}^{d}\rangle =\sum_{i}U_{\beta,i}^{d}\vert\nu_{i}\rangle,
\end{equation}
where the mixing matrices $U_{\alpha,i}^{p}$ and $U_{\beta,i}^{d}$ are connected with the NSI Hamiltonians for production and detection
\begin{equation}
\vert U_{\alpha,i}^{p}\vert^{2} \propto \vert \langle \nu_{i}; f_{P}\vert H^{P}\vert l_{\alpha}; i_{P}\rangle\vert^{2}, \ \ \ \label{new production and detection mixing matrices}  \vert U_{\beta,i}^{d}\vert^{2} \propto \vert \langle l_{\beta}; f_{D}\vert H^{D}\vert \nu_{i}; i_{D}\rangle\vert^{2}.
\end{equation}
The $i_{P(D)}$ and $f_{P(D)}$ are  initial  and  final states for production (detection) process. Then the amplitude of neutrino transition from the initial state $\vert i \rangle$  to the final state $\vert f \rangle$ (given by Eqs. (\ref{neutrin SM states}) or (\ref{new production and detection states})) is defined in the standard way 
\begin{equation}
\label{transition amplitude}
A_{f,i} = \langle f \vert e^{-iHt} \vert i \rangle,
\end{equation}
where $H$ is Hamiltonian of free neutrinos (in vacuum) or describing neutrino interaction with matter particles (in matter).
In this description of the  oscillation phenomena there is no possibility to take into account any kind of intrinsic correlation between neutrinos and other accompanying particles, but such internal correlations exist in some models of NSI.

The natural description of the oscillation phenomena is the QFT approach, where the production and detection neutrino states are not used, all calculations are done in the neutrino mass base and it is not necessary to assume that oscillations are described by  production and detection independent probabilities \cite{QFT}. In this description there is however one not very natural assumption that neutrinos propagate as virtual particles over macroscopic or even astronomical distance. Therefore another way of describing the oscillations, without the propagation of virtual neutrinos, where the QFT is used to construct the neutrino production and detection states only, has been proposed \cite{Giunti}.  Taking into account that the QM (internal wave packets) and QFT (external wave packets) descriptions of neutrino oscillations are equivalent we proposed the generalization of both approaches where any kind of NSI can be taken into account \cite{Papers on oscillation in NSI}. The proposed description can be summarized in the following points: \\

\textbf{(i)} If the $M^{P, \alpha}_{i,\lambda}([\overline{p}],[\overline{\lambda}]) $ are the amplitudes for $"\alpha"$ neutrino produced in the mass $"i"$ and helicity $"\lambda"$ state in the process: $A \rightarrow B + \overline{l}_{\alpha} + \nu_{i}(\lambda)$, and the  $M_{k,\mu}^{D, \beta}([\overline{q}] [\overline{\mu}])$ describe the  $"\beta"$  neutrino  with mass $"k"$ and helicity $"\mu"$ in the detection process  $\nu_{k}(\mu) + C \rightarrow l_{\beta} + D$  then the amplitude for the $\alpha \rightarrow \beta$ transition after time T and at a distance L  is given by \cite{Akhmedov and Kopp}
\begin{eqnarray}
\label{QFT amplitude} 
i A_{\alpha \beta}(\lambda, \mu;[\overline{p}],[\overline{\lambda}],[ \overline{q}],[\overline{\mu}]) = \hspace{6cm}  \\ \Theta(T) \sum_{k=1,2,3} M^{P, \alpha}_{k,\lambda}([\overline{p}],[\overline{\lambda}])  M_{k,\mu}^{D, \beta}([\overline{q}], [\overline{\mu}]) \int\!\! \frac{d^{3}p}{(2\pi)^{3}} F_{P,k}(E_{k}(\vec{p}\,),\vec{p}\,) F_{D,k}(E_{k}(\vec{p}\,),\vec{p}\,) e^{ -iE_{k}(\vec{p}\,) T+i\vec{p}  \vec{L}}, \nonumber
\end{eqnarray}
where the $[\overline{p}]$ and $[\overline{\lambda}]$ $\{ [\overline{q}]$ and $[\overline{\mu}]\}$ describe the average momenta and helicities of all other particles  in the production $(A,B,l_{\alpha})$  $\{$detection $(C,D,l_{\beta})\}$ process. The functions  $F_{P,k}(E_{k}(\vec{p}\,),\vec{p}\,)$ and $F_{D,k}(E_{k}(\vec{p}\,),\vec{p}\,)$ are  constructed from momenta distribution functions for all particles in the production and detection processes respectively without neutrinos (for details see \cite{Akhmedov and Kopp}) and $\Theta(T)$ is the Heaviside step function. \\ 

\textbf{(ii)} The probability for the overall process - production, oscillation and detection - is calculated as the modulus square of the amplitude (\ref{QFT amplitude}) with proper averaging over initial and sum over final particles helicities and averaging over the necessary particles momenta in the same Lorentz frame (usually the rest frame of a detector). Practically, it is more convenient to calculate the production and the detection processes separately in their own rest frames, and in the final oscillation rates to distinguish  the initial neutrino state and the final cross-section for the detection process. \\

\textbf{(ii(1))} In the rest frame of A the neutrino initial state is calculated from dynamics of the production process $A \rightarrow B + \overline{l}_{\alpha} + \nu_{i}(\lambda)$ in the well know way.  The neutrino density matrix in the mass ($i$) and helicity ($\lambda$) base $\vert i,\lambda\rangle$ is equal
\begin{eqnarray}
\label{neutrino density matrix}
\varrho ^{\alpha}_ {\lambda, i; \eta, k} ( E,\theta,\varphi) = \hspace{5cm} \\ 
 \frac{1}{N_{\alpha}} \sum_{\lambda_{A},\lambda_{A'},  \lambda_{B},\lambda_{l}}  \int \overline{dLips} M^{P,\alpha}_{i,\lambda} (\lambda_{A};\lambda_{B},\lambda_{l}; E,\theta,\varphi) \varrho_{\lambda_{A},\lambda_{A'}}  M^{P,\alpha *}_{k,\eta} (\lambda_{A'};\lambda_{B},\lambda_{l}; E,\theta,\varphi), \nonumber
\end{eqnarray}
where the integral $\overline{dLips}$ is taken over the part of the phase space, without neutrino energy (E) and its momentum direction $(\theta,\varphi)$,  the $\varrho_{\lambda_{A},\lambda'_{A}}$ describes the polarization density matrix of decaying particle (A) and the  factor $N_{\alpha}$ normalizes the density matrix, such that  $Tr \varrho  =1$. It is convenient to normalize separately the initial state and the final cross section, then for the plane wave (not wave packet) approach the final oscillation rate is already normalized properly. 

\textbf{(ii(2))} To get the density matrix in the LAB frame,  where decaying particles are moving (pions, decaying nuclei, muons),  the Lorentz transformation, equivalent to the Wigner rotation, $\varrho_{LAB} = D^{1/2}_{W} \varrho_{CM} D^{1/2 \dagger}_{W}$ has to be performed.  For relativistic neutrinos in the helicity base, the matrix $D^{1/2}_{W}\simeq 1$, then the $\rho_{LAB}$  to a very good approximation is equal to $\varrho^{CM}$ \cite{Papers on oscillation in NSI}
\begin{eqnarray}
\label{Lorentz boost}
\varrho ^{\alpha}_{LAB}( E_{b},\theta _{b},\varphi ) = D^{1/2}_{W} \varrho_{CM}^{\alpha} ( E,\theta,\varphi) D^{1/2\dagger}_{W} \approx  \varrho_{CM}^{\alpha} \equiv\varrho^{\alpha} ( E,\theta,\varphi), 
\end{eqnarray}
where  the neutrino energy ($E$) and its spherical angle ($\theta $) are expressed in terms
of the appropriate quantities $E_{b}$ and $\theta _{b}$ in the LAB frame after the Lorentz transformation.

\textbf{(ii(3))} As neutrino detector is usually far away from the production place, to all practical purpose it is possible to use average density matrix defined in the following way 
\begin{eqnarray}
\label{average ro}
\overline{\varrho}^{\alpha}(E_{b})=\frac{1}{\frac{d\Gamma(\Delta_{b})}{dE_{b}}}\int_{\Delta_{b}} d\Omega_{b} \varrho^{\alpha}(\theta_{b},\varphi) \frac{d^3\Gamma}{dE_{b}d\Omega_{b}}, \  \  \  \frac{d\Gamma(\Delta_{b})}{dE_{b}} =\int_{\Delta_{b}} d\Omega_{b}  \frac{d^3\Gamma}{dE_{b}d\Omega_{b}},
\end{eqnarray}
where $\Delta_{b}$ is the solid angle in which neutrinos "see" the detector. The $\frac{d^3\Gamma}{dE_{b}d\Omega_{b}}$ is the neutrino energy and angular distribution. For large baseline $L$, practically $\overline{\varrho}^{\alpha}(E_{b}) \simeq\varrho^{\alpha}(\theta_{b}=0,\varphi)$. \\

\textbf{(iii)} The detection cross section is calculated in the standard way. Such a cross section describes  the neutrino oscillation and  detection process in one formula. Neglecting the coherence effect connected with the particles wave packets, the oscillation cross section is given by 
\begin{eqnarray}
\label{detection cross section}
 \sigma_{\alpha \rightarrow\beta} (E_{b},L)= \hspace{6cm} \\  \frac{1}{32\pi s}\frac{|\vec{p}_{f}|}{|\vec{p}_{i}|}\frac{1}{2s_{C}+1} \sum_{i,k;\lambda,\eta; [\overline{\mu}]}\int dLips \ M_{i,\lambda}^{D,\beta}(\vec{p}_{f}, [\overline{\mu}]) \ \overline{\varrho}^{\alpha}_{i,\lambda; k,\eta}(E_{b})\, e^{i \textstyle \frac{(m_{k}^{2} -m_{i}^{2})L}{2E_{b}}} \   M_{k,\eta}^{D,\beta *}(\vec{p}_{f}, [\overline{\mu}]). \nonumber
\end{eqnarray} \\

The two formulae (\ref{neutrino density matrix}) and (\ref{detection cross section}) are crucial for our analysis. From (\ref{neutrino density matrix}), calculating $Tr((\varrho^{\alpha})^{2})$, we see when the neutrino state is pure, or when it have to be considered as mixed. For example we can easy find that in the SM,
\begin{equation}
\label{ro for SM}
(\varrho^{\alpha}_{ \lambda, i; \eta,k})^{SM} = \delta_{\lambda,\eta}  \delta_{\lambda,-1} U_{\alpha,i} U^{*}_{\alpha,k}\, ,
\end{equation}
which describes the pure quantum mechanical state. However, there are models of NSI in which neutrinos with both helicities can be produced. In such models, in the obvious way, the neutrino production  states are mixed. But even if only the relativistic neutrinos with negative helicities are produced the states can be mixed too, it depends on the specific structure of the helicity amplitudes (see e.g. \cite{Sz Z}). From the Eq. (\ref{detection cross section}) we can easy find when the final oscillation rate factorize and when not.
\\

In conclusion, we have presented the theory of neutrino oscillation applicable in the case of any NSI, where the internal degrees of freedom of neutrinos and other accompanying particles in a production and/or in a detection process are  entangled.\\

\textbf{Acknowledgements} 

This work has been supported by the Polish Ministry of Science under grant No. N  N202 064936. M.Z. would like to thank the organizers of NOW 2010 for their hospitality. R.S. acknowledges a scholarship from the UPGOW project cofinanced by the European Social Fund.

\end{document}